# Field Free Spin-Orbit Torque Controlled Synapse and Stochastic Neuron Devices for Spintronic Boltzmann Neural Networks

Aijaz H. Lone,[1] Meng Tang,[2] Camelia Florica,[3] Bin He,[4] Jingkai Xu,[§] Xixiang Zhang,[§] and Gianluca Setti[5]

Spintronics offers a promising approach to energy-efficient neuromorphic computing by integrating the functionalities of synapses and neurons within a single platform. A significant challenge, however, is achieving field-free spin–orbit torque (SOT) control over both synaptic and neuronal devices using an industry-adopted spintronic materials stack. In this study, we present field-free SOT spintronic synapses utilizing a CoFeB ferromagnetic thin film system, where asymmetrical device design and specifically added lateral notches in the CoFeB thin film facilitate effective domain wall (DW) nucleation, movement, and pinning/depinning. This method yields multiple analog, nonvolatile resistance states with enhanced linearity and symmetry, resulting in programmable and stable synaptic weights. We provide a systematic measurement approach to improve the linearity and symmetry of the synapses. Additionally, we demonstrate nanoscale magnetic tunnel junctions (MTJs) that function as SOT-driven stochastic neurons, exhibiting current-tunable, Boltzmann-like probabilistic switching behavior, which provides an intrinsic in-hardware Gibbs sampling capability. By integrating these synapses and neurons into a Boltzmann machine complemented by a classifier layer, we achieve recognition accuracies greater than 98% on the MNIST dataset and 86% on FashionMNIST. This work establishes a framework for field-free synaptic and neuronal devices, setting the stage for practical, materials-compatible, and all-spintronic neuromorphic computing hardware.

## I. INTRODUCTION

AI technologies are driving the global economy, but training and testing large neural networks face energy challenges due to issues such as CMOS scaling and the memory-CPU bottleneck [1, 2]. Neuromorphic computing, inspired by the brain's parallel, event-driven, and energy-efficient processing, has emerged as a promising alternative [3–5]. However, to realize hardware neuromorphic systems, we need devices that can replicate the functional behavior of biological neurons and synapses while ensuring low power consumption, high endurance, and scalability [5]. Significant research and technology development in the field of neuromorphic devices is underway, utilizing various memristive device concepts, including resistive memristor devices[6], phase change memory PCM [7], ferroelectric devices [8], and spintronics[9, 10]. These innovative approaches to developing neuromorphic computing architectures based on beyond-CMOS memory devices show great potential for addressing the energy challenges associated with AI training using conventional computing methods [1]. Among them, spintronic devices, which leverage the spin degree of freedom for information storage, have emerged as a sustainable data storage technology [11, 12] and a promising neuromorphic hardware primitive such as synapses, neurons, p-bits [10, 13– 16] and true random number generation [17–19]. The MTJ is regarded as a state-of-the-art spintronic memory device, featuring non-volatile data storage, endurance exceeding $10^{12}$ cycles, a tunable energy barrier, good CMOS compatibility, and sub-nanosecond read and write times, as well as digital and analog switching capabilities [20–22]. These MTJ merits also benefit spintronic-based neuromorphic hardware, making these devices a suitable fit for various neuromorphic architectures, including artificial neural networks, spiking neural networks, and probabilistic computing. Thus, spintronic devices for neuromorphic computing are emerging, as artificial counterparts of biological synapses [22–25] and neurons [26–30]. Particularly, spintronic devices based on magnetic domain walls (DWs) and skyrmions have demonstrated diverse applications in neuromorphic computing [31–40]. Among these, spin-orbit torque (SOT) devices are of particular interest because they can decouple read and write paths, enabling faster and more reliable switching [41, 42]. Using SOT-driven domain wall motion, researchers have shown synapse and neuron-like properties in multiple studies [31–36]. However, for perpendicular magnetic anisotropy material devices, achieving deterministic SOT requires breaking the symmetry of the system, typically achieved by applying an external magnetic field in the direction of the current $H_x$ [43]. This requirement can hinder the scalability of SOT-driven spintronic devices in high-density neuromorphic or data storage applications. Achieving field-free spin–orbit torque (SOT) switching has been a central challenge for scaling spintronic memory and neuromorphic devices. Various strategies have been proposed to overcome this limitation. Early work by Yu et al. demonstrated deterministic switching by introducing wedge-shaped deposition-based structural symmetry breaking in *Ta/CoFeB/TaO$_x$* [44]. Subsequent efforts employed hybrid approaches combining spin-transfer torque STT and SOT to realize field-free switching in SAS-MRAM [45, 46], or utilizing the interplay of SOT, exchange bias, and voltage-controlled anisotropy in VGSOT-pMTJ devices [47]. The first field-free perpendicular SOT-MRAM, incorporating a hybrid spin source that provides built-in symmetry breaking, was presented in [48]. Other works have leveraged exchange field gradients to induce internal effective fields for field-free switching [49]. More recently, two-dimensional van der Waals materials-based heterostructures have been used to demonstrate deterministic field-free switching above room temperature [50]. Together, these diverse approaches highlight the breadth of material and device innovations being pursued to eliminate the need for external fields in SOT technologies, a crucial step toward their practical deployment. However, the realization of field-free SOT-driven domain wall motion-based spintronic synapse demonstrating multi-state nonvolatile programmable plastic conductance states, robust linearity, and potentiation/depression symmetry becomes essential for the realization of spin-based neuromorphic hardware. In neuromorphic spintronics studies, the focus has been mainly on either independent spintronic devices as weights or spintronic devices as spiking elements, such as leaky integrate-and-fire neurons. Although spiking neural networks have shown some advantages over artificial neural networks, these networks still struggle due to a mismatch between algorithms and hardware (spiking neurons). In parallel, the stochastic SOT-driven switching of the MTJ in the absence of $H_x$, although seen as a limitation, is interesting for controlling this stochasticity in the realization of stochastic neurons. As both synapses and neurons are based on the same material stack, it is interesting to examine the functionalities of both synapses and neurons within the same spintronic system. In this work, we develop and experimentally demonstrate a field-free spin-orbit torque (SOT)-based spintronic synapse and a current-controlled MTJ-based Boltzmann stochastic neuron. The field-free SOT is achieved using an asymmetric Hall bar design, and artificial notches are added to the edges for enhanced DW pinning. Thus, apart from the linearity, the synapses also demonstrate improved retentivity. We present a pulse engineering scheme for the synapse that enables controlled and near-linear potentiation/depression behavior, critical for learning stability in neuromorphic systems. For the neuron, we achieve probabilistic switching behavior governed by the magnitude of the input current, mimicking the activation dynamics of biological neurons and facilitating stochastic neural inference. We integrate these spintronic synaptic and neuronal elements into a Boltzmann neural network neuromorphic computing framework and validate their functionality on benchmark pattern recognition tasks using the MNIST and Fashion-MNIST (FMNIST) datasets. Our results demonstrate the feasibility of using field-free SOT and MTJ-based devices as core building blocks for energy-efficient and scalable hardware neural networks. This work develops a framework for all-spintronic-based RBM hardware, mitigating the technology integration issue, and paves the path towards physics-based AI.

[1] Computer, Electrical and Mathematical Sciences and Engineering (CEMSE) Division, King Abdullah University of Science and Technology (KAUST), Saudi Arabia; Contact author: aijaz.lone@kaust.edu.sa
[2] Computer, Electrical and Mathematical Sciences and Engineering (CEMSE) Division, King Abdullah University of Science and Technology (KAUST), Saudi Arabia; Contact author: meng.tang@kaust.edu.sa
[3] Nanofabrication Core-labs, King Abdullah University of Science and Technology (KAUST), Saudi Arabia
[4] Physical Sciences and Engineering (PSE) Division, King Abdullah University of Science and Technology (KAUST), Saudi Arabia
[5] Computer, Electrical and Mathematical Sciences and Engineering (CEMSE) Division, King Abdullah University of Science and Technology (KAUST), Saudi Arabia; Contact author: gianluca.setti@kaust.edu.sa

## II. RESULTS

**Realization of the magnetic field-free SOT-controlled spintronic synapse:** Spintronic synapses are fundamental building blocks for realizing neuromorphic computing hardware. These devices enable energy-efficient analog weight storage. Fig. 1 shows the magnetic field-free spin–orbit torque (SOT)-controlled spintronic device structures. These structures are designed to operate as artificial synapses with properties including multiple non-volatile and plastic (programmable) conductance values, enabling the trainable analog weights of the hardware neural network. The material stack used for this purpose is [Ta(5)/CoFeB(0.65)/MgO(2)/Ta/TaOx(2)], where the numbers in parentheses represent the thickness in nanometers (nm). Fig. 1(a) illustrates the first device design, which consists of a 4 $\mu$m × 50 $\mu$m track with multiple lateral reading arms. We designed two types of devices:

1. Devices with an intentional channel offset design with CoFeB/MgO/Ta channel asymmetrically placed over the bottom heavy metal Ta, but no artificial notches were introduced in these devices, see Fig. 1(a-b).

2. Devices structure with multiple notches on both lateral sides of the magnetic (CoFeB) track and a slight unintentional (fabrication-induced) offset, see Fig. 1(c-d).

The SEM images corresponding to these devices reveal a clear offset for the devices shown in Fig. 1(a), as reflected in Fig. 1(b). Fig. 1(d)shows the SEM top view and its zoomed version of the notched geometry design in Fig. 1(c), where notches of size 1 $\mu$m × 1 $\mu$m can be observed. The artificial channel offset is intentionally introduced to create an asymmetrical current flow, such as a current gradient, within the CoFeB layer, which enables the field-free domain wall switching. The purpose of incorporating artificial notches is to enable controlled domain wall pinning, which facilitates field-free switching and simultaneously allows for controlled tunability of synaptic resistance modulation and enhanced retention of conductance states. The combined effects of the channel offset, multiple reading arms, and notches enable the realization of field-free spin-orbit torque (SOT) driven DW motion. Fig. 1(e) shows the weight plasticity characteristics obtained in these synaptic devices, and Fig. 1(f) illustrates the stochastic MTJ neuron device and its integration with spintronic synapses for the realization of multiply accumulate and sampling operation, typical to stochastic neural network functioning. Fig. 2 shows the SOT-driven switching results obtained in three different types of devices with $H_x$=0. Fig. 2(a) shows the anomalous Hall resistance vs. external perpendicular magnetic field Hz for three device designs (1). The hysterisis, shown in blue, corresponds to the device with artificial notches and a slight unintentional offset (corresponding to Fig. 1(a)). The hysteresis of the asymmetric etched device with CoFeB/MgO/Ta layers placed on the bottom Ta with a lateral down offset (OD) and no artificial notches is shown by red, and the results in green correspond to the RH characteristics of the device with no notches but a lateral upward offset (OU). The $RH_z$ characteristics of these devices indicate that the devices possess a reasonable perpendicular magnetic anisotropy (PMA), which is the first crucial step towards realizing spintronic synapses based on SOT-controlled domain wall (DW) motion. Fig. 2(b) shows the current (SOT) controlled device characteristics in these devices. All these SOT measurements were carried out in the unloaded position, which means no external magnetic field was involved in the measurement. To obtain the IR hysteresis loop, the current was swept from -3 mA to +3 mA and vice versa. We observed partial field-free SOT-driven resistance switching in a device labeled (NO), having artificial notches (NO), but no offset. The devices labeled (OU) with upward offset and no notches showed no field-free SOT-driven switching. In contrast to the other two devices, the device labeled (OD) with a downwards offset shows reliable field-free switching, as indicated by the red hysteresis loop in Fig. 2(b). The observation of field-free SOT switching in notched and down offset devices is attributed to the following reasons. Artificial notches locally break symmetry and introduce non-uniform demagnetizing fields, creating a magnetization tilt that mimics an in-plane bias. Asymmetric positioning of the CoFeB/MgO stack above the heavy metal Ta breaks structural inversion symmetry, leading to an imbalanced spin current distribution and net effective SOT. The interfacial Dzyaloshinskii-Moriya interaction (DMI) further stabilizes fixed-chirality N´eel domain walls, thereby enhancing deterministic switching in the device (OD) only due to chirality. Notches also induce localized Oersted fields due to current crowding, contributing additional adequate in-plane torque. These notches also act as pinning sites to trap domain walls, which improves deterministic switching. The Non-uniform heating during current flow introduces thermal gradients, potentially generating spin Seebeck-like torques that assist switching. Finally, asymmetric geometry causes non-uniform spin-Hall current injection, resulting in a spatially varying torque profile with a net in-plane component. Collectively, these effects break symmetry and enable reliable field-free magnetization switching. We quantify the degree of field-free SOT switching by defining the bit error rate BER. We applied alternating positive and negative current pulses of amplitude (+3 mA, -3.5 mA) and width 100 $\mu$s to set the device into $R_H$ and reset it into $R_L$, repeating this process. After each writing event, device resistance was measured by applying a read current pulse of 0.1 mA. The writing error or failed event is noted whenever the measured resistance change is less than 25% of the entire memory window ($R_H - R_L$). BER is defined as the number of measured writing events that failed to switch resistance beyond 25% of the total $R_{max}$. As shown in Fig. 2(c) for the Hall bar without any offset, a BER of 6% was measured (see Fig. 2(d)). If a stricter bar is set for noting failed events, we expect this percentage to increase. Probed RH means the measured anomalous Hall resistance after each writing event. In contrast, the devices with down offset exhibited nearly ideal field-free SOT switching, as shown in Fig. 2(e). We noticed negligible switching resistance variation and only two failed writing events were recorded, resulting in 2% BER as demonstrated in Fig. 2(f). Clearly, even with an increased bar defining a successful writing event, these devices will still show very reduced BER due to complete resistance switching.

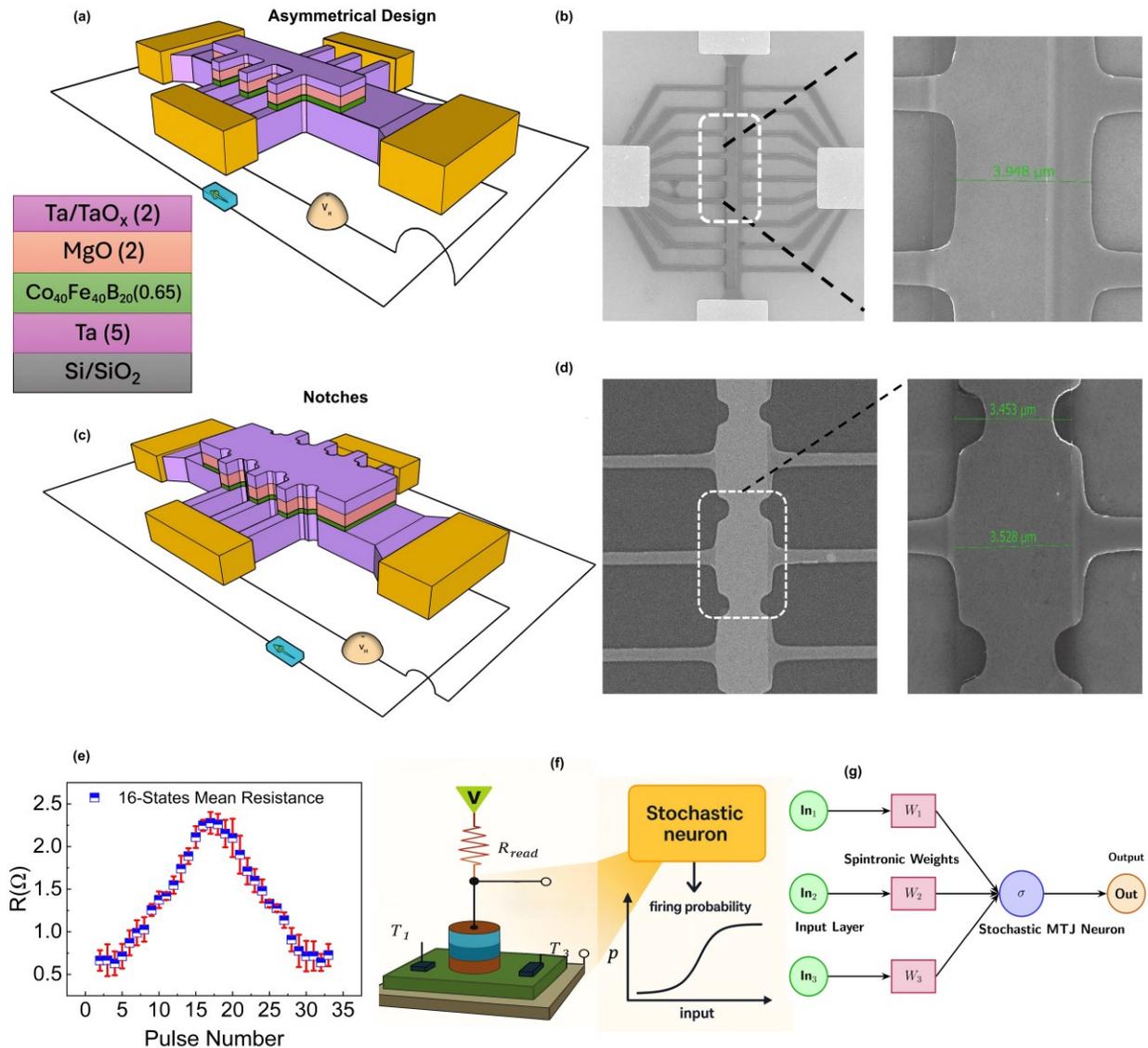

Figure 1. Field-free SOT domain wall synapse:(a) Multi-arm Hall bar device structure with intentional offset design geometry, (b) SEM image showing the top view of the device, and its zoomed-in region clearly shows asymmetric etched geometry where the CoFeB/MgO/Ta channel is asymmetrically placed over heavy metal Ta. (c) Multi-arm Hall bar device structure with notches; the CoFeB/MgO/Ta channel is symmetrically placed over heavy metal Ta, but there exists a slight unintentional offset in the design. (d) Scanning electron microscopy image of the device shown in (c), and its zoomed-in version. (e) Weight plasticity characteristics of the synapse and (f) Stochastic MTJ neuron and its integration with synapses for the realization of multiply accumulate and sampling operation.

To obtain the linearly programmed resistance, we measured two types of devices. (1) Single readout: In this type of device, the transverse Hallbar arms are separated from each other, so reading was done by connecting one of the arms to the nano-voltmeter. We observed a sharp transition in resistance for this case because, in anomalous Hall measurements, the $R_{XY}$ depends on the magnetization centered around the transverse arm, which in this case switches sharply. (2) Combined: In this device, the multiple Hall arms are shorted, and resistance is measured. The anomalous Hall resistance is given by

$$\rho_{xy} = R_O B + R_S \mu_0 M_Z + R_O P B_{em}^z \tag{1}$$

In most cases, the resistance depends on the local $M_z$ component of the magnetization, which is centered between the two measuring arms. In the single-arm readout scheme, the switching is abrupt; however, in a multi-arm setup, the magnetization change measured by each arm is accumulated as domains expand. This results in gradual resistance switching as noticed in Fig. 3(a). Fig. 3(b) shows the resistance switching (potentiation/depression) cycles at different $H_x$. The potentiation/depression cycle was repeated five times; the average of these five cycles for each $H_x$ is shown. The multiple analog resistance states clearly reflect the SOT-controlled DW motion or domain expansion and contraction. In the creep region, the domain wall velocity depends on the local pinning potential, the critical depinning field, the current, and the external magnetic field. A current in the heavy metal generates a damping-like SOT that drives Néel-domain-wall motion in the CoFeB layer. DMI fixes the DW chirality (internal angle $\phi$). An in-plane field $H_x$ sets DW angle $\phi$ and thus increases the SOT efficiency $\propto \cos\phi$. The effective drive $H_{eff}(J, H_x)$ enters the DW velocity; in the creep/near-threshold region

$$v \approx v_0 \exp\left[-\left(\frac{H_c}{H_{\text{eff}}}\right)^{\mu}\right], \quad \mu \approx 1/4 \tag{2}$$

So at fixed current amplitude, the increasing $H_{\text{eff}}$ (via $+H_x$ aligned with chirality) rapidly increases $v$. This lowers the depinning potential, reduces Barkhausen jumps, and yields fewer resistance states, resulting in more abrupt updates as seen in Fig. 3(b) and presented in Fig. 3(d). Around the zero $H_x$, the DW encounters intermittent pinning potentials which result in increased discrete resistance states and improved linearity. We model each branch ($n$ = 1...32 pulses) by a logistic model, where the resistance $R_{pot}$ and $R_{dep}$ are non-linear functions of the pulse number, and asymmetry is captured by a separate field-dependent slope parameter $k(H_x)$ for potentiation and depression:

$$R_{\text{pot}}(n) = R_{\min} + \frac{\Delta R}{1 + \exp[-k(H_x)(n - n_0)]}, \tag{3}$$

$$R_{\text{dep}}(n) = R_{\max} - \frac{\Delta R}{1 + \exp[+k(H_x)(n - n_0)]} \tag{4}$$

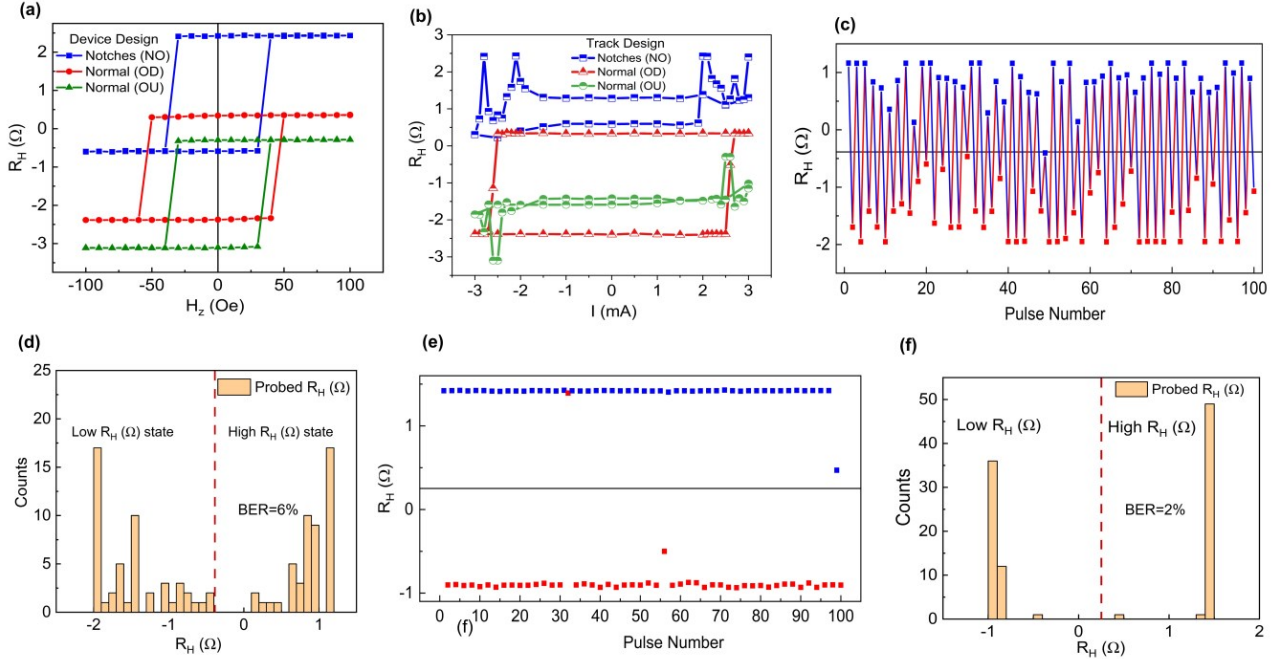

Figure 2. Quantification of field-free switching: (a) RH hysteresis characteristics in three device geometries of Notches but no offset (NO), Normal (offset down), and Normal (offset up). (b) Field-free SOT-driven switching in three geometries clearly shows the strongest deterministic switching in the offset-down sample; the notched device without offset exhibits a reduced switching window around 0 Oe due to domain wall pinning, and no switching occurs in the offset-down sample. (c-d) SOT-driven switching events in NO(Notched device) show a BER OF about 6%, and (e-f) in an offset down asymmetric device, the BER is reduced to 2%.

In terms of magnetization evolution, the sigmoid characteristics arise from (i) nucleation + growth $df/dn = \alpha f(1-f)$ or (ii) thermally assisted multi-site switching with a per-pulse success probability $p$ over a finite window. Each measured curve has its local logistic fit, and we fit a global logistic function corresponding to all these results, represented by a continuous black line shown in Fig. 3(b). The slope of the potentiation and depression curves is computed by:

$$S \equiv \left.\frac{dR_H}{dn}\right|_{n_0} = \frac{k\,\Delta R}{4} = A\kappa. \tag{5}$$

As $H_x$ increases from (0 to 6) Oe, the pulse vs slope (S) profiles for potentiation (P) and depression (D) characteristics corresponding to these field values become taller and narrower: the mid-range per-pulse update $S_{max} = k\,\Delta R/4$ grows while the transition spans fewer pulses (larger $k$). Physically, $H_x$ sets the domain-wall angle along with DMI-fixed DW chirality, enhancing the effective SOT drive $H_{eff} \propto H_{DL}\cos\phi$ and the DW velocity in the creep regime, yielding stronger, less Barkhausen-like updates. P and D differ because chirality allows one direction to couple more efficiently with SOT. Near the DMI-compensation field, the two bells converge, and symmetry improves $(S_{max}^P \approx |S_{max}^D|)$. The trade-off for global linearity is that very large $k$ concentrates updates into a few pulses (early saturation), while very small $k$ spreads them too weakly; optimal linearity occurs when the 10–90% transition spans ∼ the 32-pulse branch.

We performed additional measurements on the characteristics of potentiation and depression at zero field and across various current amplitudes, as shown in Fig. 3(e). As the write current is increased, both experimental findings and micromagnetic simulations (see Fig. 3(f)) demonstrate a three-regime behavior for spin-orbit-torque (SOT)-driven domain walls within a pinned track: (i) In the low current regime (creep), minor increases in the effective SOT field $H_{DL} \propto J$ result in considerable increases in the velocity of the domain wall. As a result, the effects of depression and potentiation become more pronounced, and the plateaus start to diminish. (ii) In the intermediate current regime, the domain wall interacts significantly with geometric pinning sites (like artificial notches and pinning sites near track and read arm junctions) (see supplementary material Fig.S2 for micromagnetic images). It can tilt or precess, becoming temporarily trapped. This interaction leads to a non-monotonic dip seen in the simulated $M_z(J)$ and, in experimental observations, a lower mid-slope $S = k\Delta R/4$, a later inflection point $n_0$, and an increase in plateau counts and variance (e.g., around 2.50–2.54 mA). (iii) At elevated currents, the driving force surpasses the pinning sites, causing the motion to transition to a flow regime, thereby restoring smooth and deterministic updates. This direct correlation establishes a link between the observed synaptic nonlinearity and the SOT–DW interaction within the designed pinning arrangement of the device.

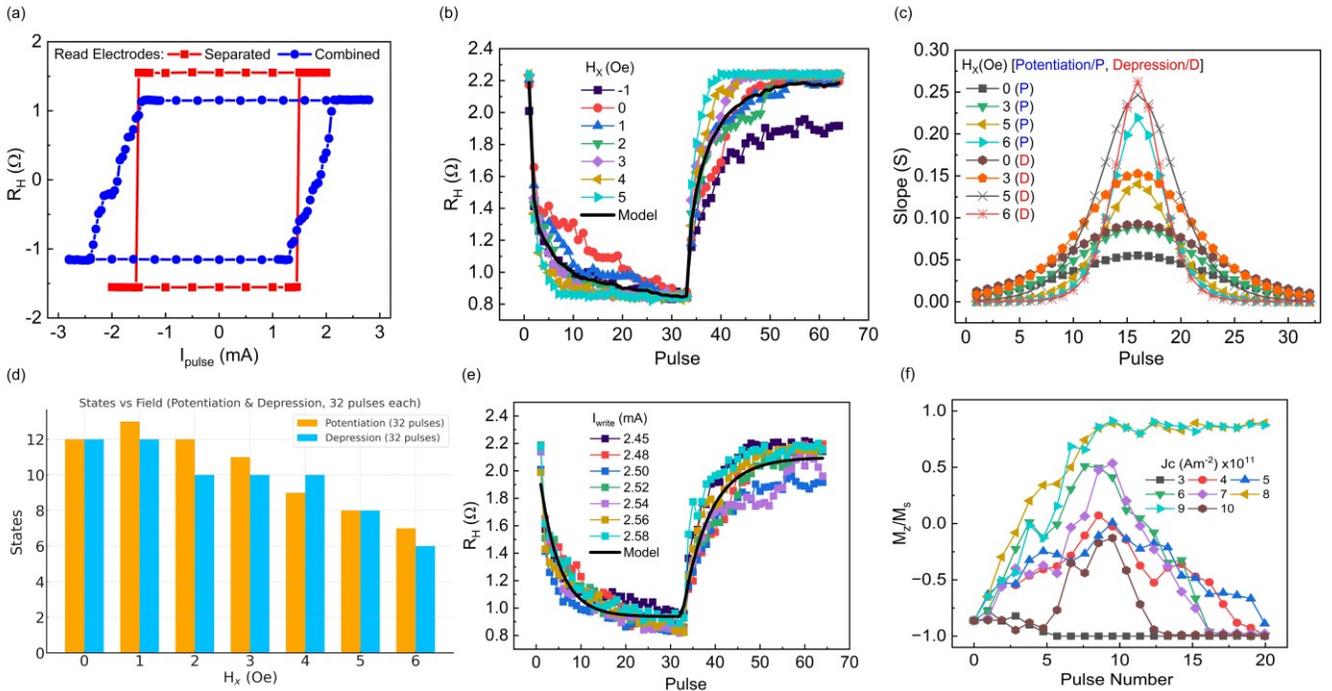

Figure 3. Field-free SOT-controlled device plasticity: (a) RI hysterisis characteristics of separate vs. combined readout showing sharp vs. gradual resistance transitions. (b) Measured (mean cycle-to-cycle resistance) potentiation/depression at $H_X$ = [−1, 0, 1, 2, 3, 4, and 5]Oe, the continuous black curve shows the $H_X$ fitted model. [Note: In our measurement, −1Oe corresponds to true 0Oe. (c) Potentiation/depression slope S at different $H_X$. (d) Number of discrete resistance states in potentiation/depression at increasing $H_X$. (e) Measured (cycle to cycle mean) resistance potentiation/depression at increasing write current pulse amplitude and fixed pulse width 50 μs, the continuous black curve shows the current-dependent resistance evolution model. (f) Micromagnetic simulations of the scaled version of the asymmetric device ($M_Z/M_S$) at increasing writing current densities show random domain wall pinning with increasing current, thus supporting the measured current-dependent resistance switching.

The magneto-optical Kerr microscopy MOKE imaging was done further to observe the field-free SOT-induced domain propagation in the notched device. To correlate the resistance evolution with MOKE imaging, we first applied 16 depression and 16 potentiation pulses and measured the resistance as shown in Fig. 4(a). The device was observed under the MOKE, and the same write pulse scheme was employed. Fig. 4(b) shows the whole device MOKE image at the beginning (1st depression pulse). The zoomed-in images in the depression phase, as shown in Fig. 4(c), clearly show bright domains expanding and covering the entire track as the current reaches -8.5 mA. This is clearly reflected in the observation of a similar trend, as shown in Fig. 4(a). Likewise, the magnetization evolution during the potentiation phase is captured by MOKE images shown in Fig. 4(d). Upon reversing the current direction and gradually increasing the pulse amplitude, we observe the dark domains starting to dominate and cover the entire track as the current reaches 8.5 mA. These images demonstrate the achievement of field-free SOT-induced domain wall motion, and hence, the device's depression and potentiation.

Fig. 5(a) illustrates that decreasing the pulse width from 100 $\mu$s to 30 $\mu$s leads to an increase in the number of discrete programmable non-volatile resistance states, rising from about 9 states at ($\tau_p = 100\ \mu s$) to 13 states at ($\tau_p$ = 10 $\mu$s). This clearly agrees with spin-orbit torque (SOT) induced domain-wall (DW) motion in a constrained track, where the displacement per pulse is scaled as follows:

$$\Delta x \approx v(J)\,\tau_p, \qquad v(J) \approx v_0 \exp\left[-\left(H_c/H_{\text{eff}}(J)\right)^\mu\right], \quad H_{\text{eff}} \propto H_{\text{DL}} \propto J, \tag{6}$$

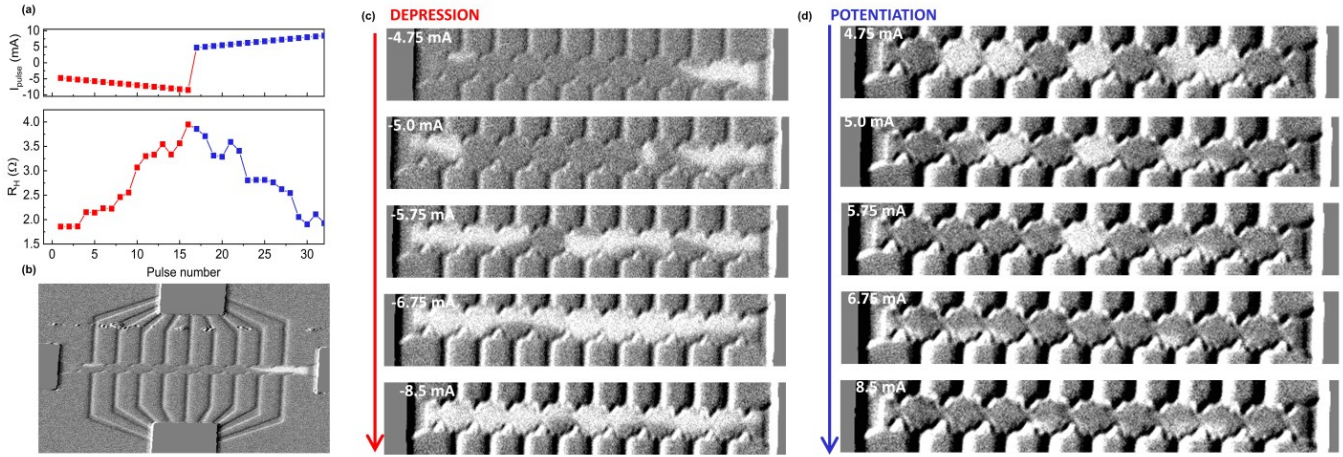

Figure 4. MOKE imaging: (a) Real-time resistance evolution with application of increasing current amplitude pulses. (b) The full MOKE image of the device under measurement reveals the nucleation of a reversed domain (bright region). (c) Zoomed-in MOKE images of the device (NO) notched during the depression phase at increasing absolute current amplitude show domain propagation and full switching, and (d) Zoomed-in MOKE images of the device during the potentiation phase at increasing current amplitude show nucleation of a reversed domain and its propagation across the track.

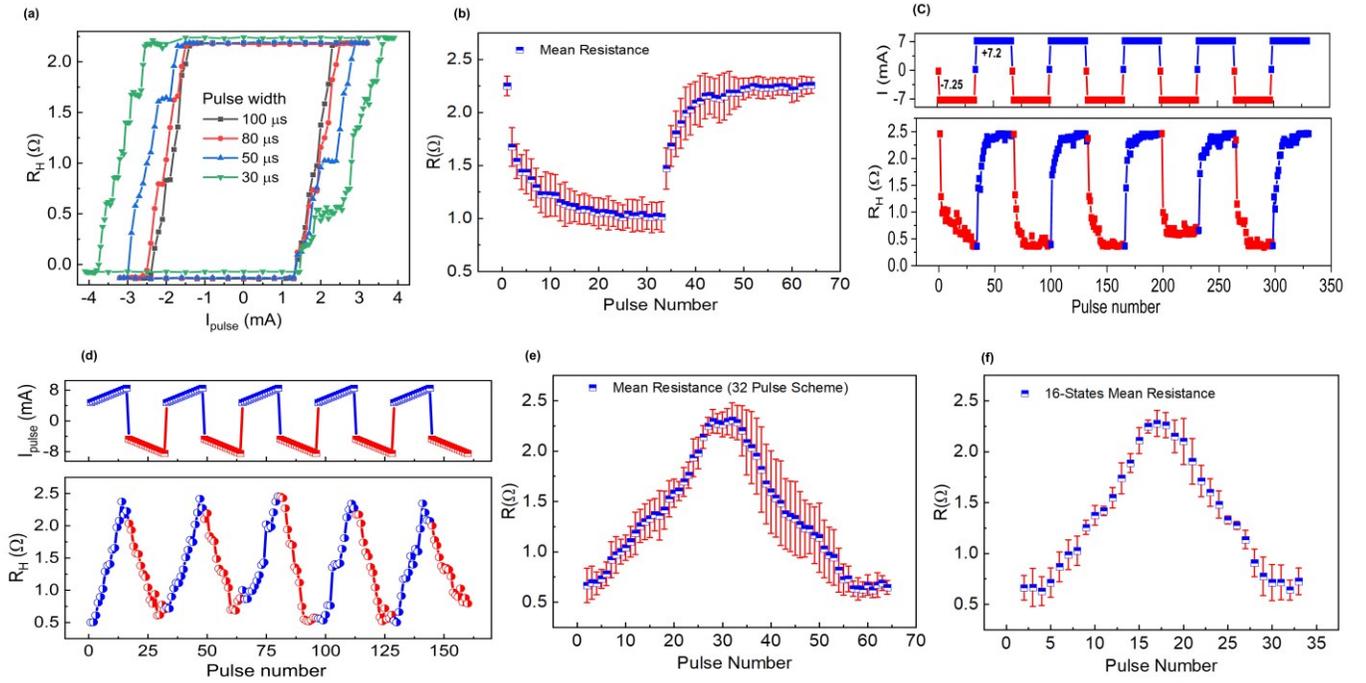

Figure 5. Tuning the linearity and symmetry of the synapse: (a) Write current pulse width dependence of the field-free SOT driven resistance switching at 30 $\mu$s, 50 $\mu$s, 80 $\mu$s, and 100 $\mu$s, shows strong discretizations (more gradual) resistance switching with decreased pulse width. (b) Mean and standard deviation of potentiation/depression at pulse width 50 $\mu$s and fixed writing current $I_n$ = -2.55 mA; $I_p$ = +2.58 mA, measured 20 cycles. (c) On application of 10 $\mu$s and $I_n$ = -7.25 mA; $I_p$ = +7.2 mA, the cycle-to-cycle variation is reduced and linearity improves slightly. (d-f) To enhance the linearity and symmetry in potentiation/depression, we apply the gradually increasing linear current ramp. (d) Repeated depression/potentiation cycles with 16-pulse scheme (e) Mean and standard deviation of linearly programmed devices with 32 pulse scheme, and (f) The linearity and symmetry are significantly increased for 16 pulse scheme, the cycle-to-cycle variation is also reduced considerably.

Long current pulses ($\tau_p$ = 100 $\mu$s) cause proportional domain wall positional changes ($\Delta x$) by overcoming several pinning sites within a single pulse, resulting in an abrupt resistance change and a reduced number of states. On the other hand, short pulses (10 $\mu$s) limit $\Delta x$ and permit a more precise sampling of the pinning landscape, thereby increasing the number of states. This indicates that even shorter pulses can further improve the number of resolvable states, linearity, and symmetry. In Fig. 5(b), the mean and standard deviation of potentiation/depression measured at fixed amplitude write current are shown. We apply ±(2.55–2.58) *mA* with 50 *$\mu$s* pulses, which produce multilevel potentiation and depression across 20 cycles, showing visible plateaus and cycle-to-cycle variation—indicating stochastic pinning and depinning. When we increase the write current amplitude to ±(7.2, 7.25) *mA while reducing the pulse width to 10 $\mu$s (as* shown in Fig. 5(c)), the device shifts away from the near-threshold creep regime. The increased spin-orbit torque (SOT) effective field $H_{\text{DL}} \propto J$ lowers the pinning energy barriers, suppresses noise, and results in more deterministic domain wall (DW) propagation. As $\tau_p$ decreases, the DW wall moves gradually; the overall result is enhanced stability, improved linearity, and symmetry between potentiation and depression. The per-branch response can be accurately described by a logistic function with an update rate *k*. Its mid-slope, $S_{\max} = k\Delta R/4$, influences the local linearity. Fixed amplitude pulses focus updates near the inflection point but can saturate at the extremes. To address this, we employed a varying-current amplitude scheme, illustrated in Fig. 5(d), where the current amplitude is linearly increased from 4.75 mA to 8.625 mA for depression and -4.75 mA to -8.625 mA for potentiation. The increasing current amplitude gradually reduces the depinning potential in proportion, allowing for a gradual movement of the domain wall due to multiple pinning events. Thus, in pulse drive, the logistic slope, which is minimal (near the extremes) and decreasing near the center, effectively flattens due to the per-pulse increment, which enhances the global linearity across the entire 32-pulse branch. Fig. 5(e-f) displays the mean and standard deviation for the 32 and 16 linearly varying pulse schemes. In our experiment, we applied 32 depression pulses ranging from 4.75 *mA* to 8.65 *mA* to nucleate and gradually expand the domain. This was followed by 32 potentiation pulses, ranging from −4.75 mA to −8.65 mA, which initiated nucleation of the reverse domain and completed the cycle. When the current is reversed, the opposite domain begins to nucleate, facilitating the overall process. We repeated this measurement 20 times to calculate the mean and standard deviation $\sigma$ of the resistance switching. This deterministic process of nucleation and propagation correlates directly with the transport data: higher currents decrease the effective pinning potential, while shorter pulses prevent overshoot. Together, these factors provide more available states, improve symmetry between potentiation and depression, and create a more linear and repeatable synaptic trajectory. However, the potentiation and depression exhibited a larger cycle-to-cycle *standard deviation (σ)* in the 32-pulse scheme. This is because utilizing 32 pulses brings domain switching closer to the threshold region, where stochasticity increases. The initial few potentiation pulses can randomly re-nucleate or de-pin from reversed domains, and the later pulses approach saturation, where spin-orbit torque (SOT)-driven domain wall motion reverts to creep, leading to minute resistance fluctuations predominating. Additionally, the extended sequence can accumulate thermal drift, resulting in more opportunities for trapping and release at defects, which inflates the standard deviation $\sigma$. In contrast, the 16-pulse protocol limits updates to the deterministic mid-slope region, where the effective drive, Heff, is larger and domain wall motion is closer to being continuous. This configuration results in a significantly smaller *standard deviation (σ)*.

### III. STOCHASTIC MAGNETIC TUNNEL JUNCTION NEURON

To realize a complete spintronic RBM, we fabricated and characterized field-free SOT-driven magnetic tunnel junction (MTJ) devices at two scales: micro (1 $\mu$m pillar) and nano (100 nm pillar). Fig. 6(a) shows the MTJ stack (thicknesses in nm): *Si/SiO$_2$/Ta*(5)*/CFB*(0.65)*/MgO*(2)*/CFB*(1.3)*/Ta*(0.6)*/Co*(0.6)*/Pt*(1.5)/ Co(0.4)/Pt(0.2) 4*/Co*(0.5)*/Ru*(1.0)*/Co*(0.5)*/[Pt*(0.2)*/Co*(0.4)]$_9$*/Ta*(8)*/Ru*(5). Here, the Ta underlayer 5 nm acts as the heavy-metal source of spin–orbit torque to switch a low-barrier CoFeB free layer 0.65 nm) through the MgO tunnel barrier 2 nm), while a CoFeB reference layer 1.3 nm is exchange-pinned by a synthetic antiferromagnet based on (Co/Pt) multilayers separated by Ru. Devices were patterned as circular pillars of nominal diameters 1 $\mu$m and 100 nm. Fig. 6(b-c) shows the field vs resistance hysteresis of the nano-MTJ and micro-MTJ devices at increasing reading current. In nano-MTJ, we achieved a tunnel magnetoresistance TMR of 33%, and in micro-MTJ, the TMR was about 44%. Considering the typical voltage bias dependence, the TMR, which shows a decrease in TMR as the voltage across the MTJ increases, we also observe a decreasing trend in both nano- and micro-MTJs with increasing read current. Furthermore, we expect TMR to improve if the read current is lowered further. Fig. 6(d) shows the SOT-driven stochastic switching in.

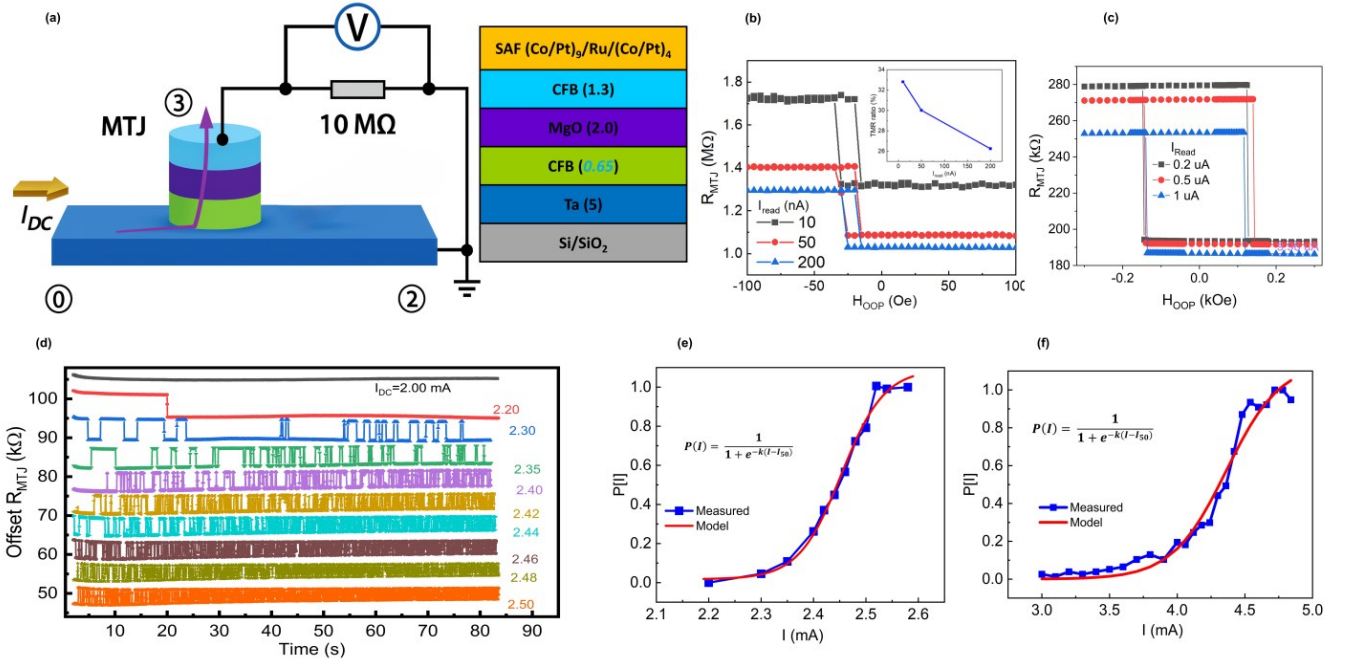

Figure 6. Realization of the stochastic MTJ neuron: (a) MTJ stochastic neuron device structure and the measurement schematic. The diameters of the MTJs are 1 $\mu$ and 100 nm. (b) The RH characteristics of a 100 nm device at increasing read currents show good PMA devices, but reduced TMR at higher read currents due to the well-known bias voltage dependence of the TMR. (c) The RH characteristics of a 1 $\mu$m device at increasing read currents. (d) SOT-controlled stochastic switching events measured in 1 $\mu$m and 100 nm [supplementary material] devices. (e-f) The current-controlled switching probability [normalized spike rate] $P(I)$ in these devices follows the Boltzmann sigmoid activation function, enabling the MTJ-based Gibbs sampling (discussed in the RBM section).

the micro-MTJ; exactly similar characteristics are measured in the nano-MTJ (Supplementary file). At zero external field in both types of devices, we observe an increasing rate of spiking (switching) from the low to the high resistance state. As seen in Fig. 6(d), at a write current of 2 mA, the MTJ remains silent; however, when the current is increased, the MTJ starts switching, albeit stochastically. In all measurements, a DC is flowing through the Ta while the MTJ resistance is measured with the nano-voltmeter (more details in the Methods section). Considering the small thickness of the Free layer (0.65 nm), the current-driven stochastic behavior observed in these devices arises from the interplay between spin-orbit torque (SOT) and thermal activation. A charge current through the Ta underlayer generates a transverse spin accumulation via the spin Hall effect, exerting torque on the ultrathin CoFeB free layer. At low current, the torque is insufficient to overcome the magnetic energy barrier, resulting in negligible switching probability P[I]. As the current increases, the P[I] also ramps up. As I approachs the critical value $I_{50}$, the SOT lowers the energy barrier, allowing thermal fluctuations to stochastically nucleate reversed domains and lead to probabilistic switching. At higher currents, the barrier collapses, and the number of stochastic switching events increases proportionally. The stochastic switching of field-free SOT-MTJs can be modeled as thermally activated domain-wall nucleation and propagation under a current-driven spin–orbit torque. The effective energy barrier is reduced both by spin torque and Joule heating,

$$\Delta E(I) = \Delta E_0 \left(1 - \frac{I}{I_c}\right) - \alpha I^2, \tag{7}$$

While a current-dependent stray field, which we observed, modifies the coercive field and moves the hysteresis loop away from the center (0 Oe),

$$H_{\text{eff}}(I) = H_c - \beta I. \tag{8}$$

The switching probability per pulse follows an Arrhenius law,

$$P(I) = 1 - \exp\left[-f_0 t_p \exp\left(-\frac{\Delta E(I)}{k_B T}\right)\right], \tag{9}$$

Which can be well approximated by a logistic function,

$$P(I) = \frac{1}{1 + \exp\left[-k(I - I_{50}(T))\right]}, \tag{10}$$

where $I_{50}(T)$ shifts with both temperature and stray field. This framework explains the measured sigmoid-like event distributions: at low current, the energy barrier is significant and $P(I) \approx 0$, while at high current, heating reduces coercivity and the stray field shifts the hysteresis, enhancing stochastic events without reaching fully deterministic switching.

### A. Neuron Model

The current-controlled stochastic switching in these devices enables their application in stochastic computing or Boltzmann machines. We model the switching probability (normalized spike rate) as the logistic function of the write current through heavy metal Ta. In both nano- and micro-MTJs, the model remains essentially the same, differing only in slope and critical current values. The neuron model for nano-MTJ devices is

$$P[I] = \frac{1}{1 + e^{-K(I - I_c)}} \tag{11}$$

where: $P[I]$ is the normalized spike probability (y–axis), $I$ is the input current in mA (x–axis). Where the fitted parameters for the nano-MTJ neuron device are: $k = 5.0$, $I_c = 4.36$ mA. For Micro-MTJ, the neuron model parameters are $k = 24.0$, $I_c = 2.44$ mA.

## IV. INTEGRATION IN RESTRICTED BOLTZMANN MACHINE

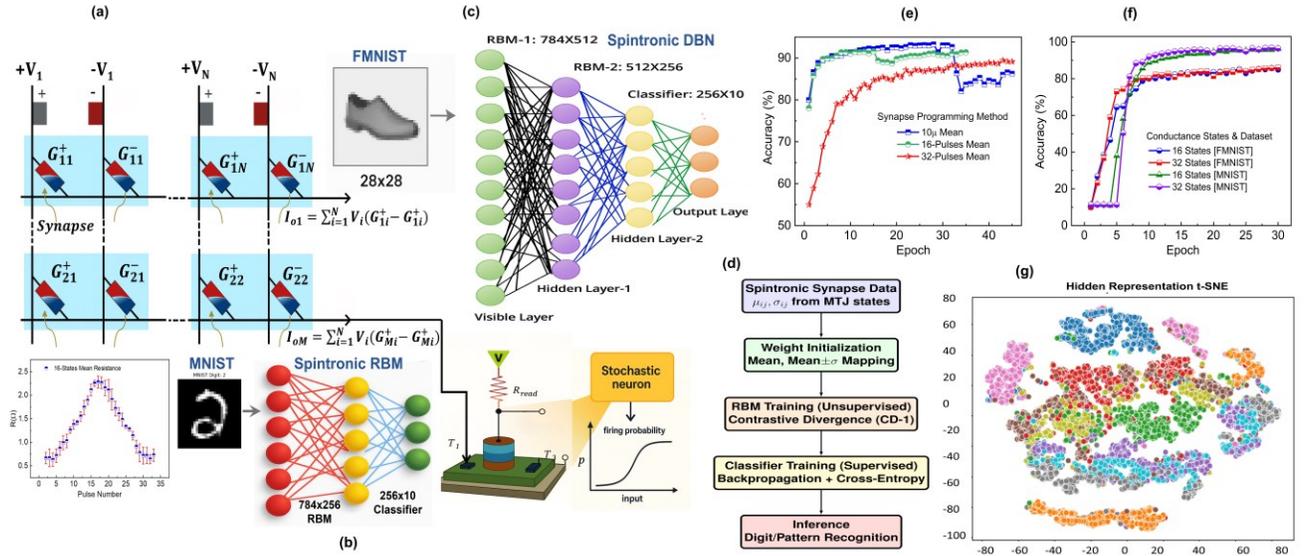

Figure 7. Spintronic synapses and stochastic neuron integration in RBM and DBN architectures: (a) spintronic synapse array in differential pair geometry for the realization of weights in the range [-1 to 1]. (b) 1-RBM + classifier architecture: 784 visible, 256 hidden, 10 output neurons for MNIST data classification. (c) DBN: 2-RBMs (784×512 and 512×256) and classifier (256×10) for FMNIST data-classification (d) RBM and classifier training and inference methodology. (e) Accuracy with epochs on the MNIST dataset for 10 μs, 16-pulse linearly increasing, and 32-pulse linearly increasing programming schemes. (f) Accuracy with epochs on the FMNIST dataset for 10 μs, 16-pulse linearly increasing, and 32-pulse linearly increasing programming schemes. (g) t-distributed Stochastic Neighbor Embedding( t-SNE of RBM shows well-separated clusters, meaning RBM learned hidden features.

The various multiple programmable quasi-linear conductance states within the spintronic synapse, along with the stochastic MTJ neurons, create an ideal foundation for developing a suitable platform for building the RBMs and deep belief networks applicable in areas such as diverse applications ranging from pattern recognition to combinatorial optimization, including tasks such as the traveling salesman problem and graph coloring. In the RBM simulations, we incorporated these spintronic synapses as symmetric weights in the RBM by employing the differential pair method (see Fig. 7(a)), enabling the realization of an approach that allows for both positive and negative weights. The MTJ stochastic neurons inherently support the in-hardware Gibbs sampling capability necessary for the RBM and DBN. We employed these MTJ devices as neurons within the RBM (784 × 256) and classifier (256 × 10) architectures for recognizing MNIST data [51], as shown in Fig. 7(b). For the FMNIST dataset[52], we scaled the network to a deep belief network DBN with RBM-1 (784 × 512), RBM-B (512 × 256), and a classifier (256 × 10) (see Fig. 7(C)). The training process in our framework utilizes a two-stage approach that combines device physics with network learning. In the RBM + Classifier illustrated in Fig. 7(b), the RBM layers undergo training using the Contrastive Divergence (CD-1) algorithm, which captures unsupervised feature representations. Following this, a softmax classifier is then fine-tuned using cross-entropy loss through backpropagation. To incorporate hardware awareness, the weights are initialized with the weights with measured MTJ synaptic states (mean, mean + $\sigma$, mean + $\sigma$), thus integrating by embedding device variability into the learning process (as shown in Fig. 7(d)). In the case of DBN, the variability from cycle to cycle is incorporated as noise. The layers of variability are added as noise. Restricted Boltzmann Machine (RBM) layers are first pretrained using contrastive divergence (CD-1), where the synaptic weights are obtained from measured device conductance states ($\mu, \sigma$), which directly integrates device variability into the learning algorithm:

$$\Delta W_{ij} = \eta\left(\langle v_i h_j \rangle_{\text{data}} - \langle v_i h_j \rangle_{\text{model}}\right) + \sigma_{ij}\, \xi_{ij}. \tag{12}$$

Fine-tuning with backpropagation in the classifier stage adapts these hardware-aware weights for task-level performance through the cross-entropy loss.

$$\hat{y} = \text{softmax}(W^{(L)} h^{(L-1)} + b^{(L)}). \tag{13}$$

We used weights derived from writing schemes that included 10 $\mu$s constant current pulses, a linearly increasing 16-pulse scheme, and a linearly increasing 32-pulse scheme. As shown in Fig. 7(e), we achieved 90% recognition accuracy for the 16-pulse scheme, which is attributed to very low $\sigma$ and the highest linearity and symmetry in this case. In a 32-pulse linearly increasing current scheme, the accuracy is approximately 89%. In contrast, the 10 μs constant current pulses scheme initially achieves an accuracy of about 93%, but ultimately settles at around 86%. In DBN, the MNIST data recognition accuracy settles around 98% for both 16 and 32 states. The FMNIST recognition achieves up to 85% accuracy in the 16-state scheme and 86% in the 32-state case (see Fig. 7(f)). The extended accuracy results for different pulse schemes are further presented in the supplementary file Table S1. Lastly, because both synapses and neurons incorporate stochasticity from MTJ physics via noisy weight sampling and probabilistic switching, the RBM naturally exhibits Neural Sampling Machine (NSM) properties, enabling inference and sampling consistent with the underlying spintronic hardware.

## V.  CONCLUSION

In summary, we demonstrate field-free, spin-orbit torque (SOT) programmed spintronic synapses and SOT-controlled magnetic tunnel junction (MTJ) stochastic neurons using the industry-standard Ta/CoFeB/MgO material stack. The development of field-free SOT neuromorphic devices is essential for scaling these technologies into neuromorphic circuits. Our device design features an asymmetric Hall bar with variations, including notches and multiple arms. The mechanism for field-free magnetization switching is based on the combined effects of artificial notches, structural inversion asymmetry, interfacial Dzyaloshinskii-Moriya interaction (DMI), and current-induced effects. Together, these components disrupt symmetry and create a robust in-plane torque, allowing for deterministic and energy-efficient magnetization control. This highlights a scalable pathway toward next-generation spintronic devices. Our design incorporates domain wall (DW) pinning, which enables current-driven DW motion and achieves multiple stable, programmable resistance states with excellent linearity and repeatability. We not only demonstrate multiple non-volatile, programmable weights in the synaptic device but also conduct an exhaustive analysis to enhance linearity, cycle-to-cycle variation, and symmetry in potentiation/depression cycles. These metrics are critical for the training and inference of neural networks in any memristive synaptic device. We measured the mean potentiation and depression resistance, along with their variance, across cycle-to-cycle programming. The MTJ neuron devices, scaled down to a diameter of 100 nm, exhibit SOT-controlled, Boltzmann-like stochastic switching characteristics. The switching probability as a function of current is modeled as a logistic neuron that performs in-hardware Gibbs sampling, making these neurons ideal candidates for restricted Boltzmann machine (RBM) architectures. We integrated these neurons as Gibbs sampling units and classifier neurons within RBM and deep belief network (DBN) architectures. The spintronic synaptic weights serve as the bipolar weights required for RBM training, enabling the capture of both positive and negative features in data. Using synaptic weights derived from various programming schemes, we achieved recognition accuracy up to 90% with the 16-pulse protocol, approximately 89% with the 32-pulse protocol, and around 86% with constant pulses. In DBN, the classification accuracy reached 98% on MNIST and 85–86% on FMNIST. The intrinsic stochasticity of the MTJ synapses and neurons further enables the RBM properties akin to a Neural Sampling Machine, allowing direct probabilistic inference mapped to spintronic hardware. Our demonstration of field-free domain-wall synapses on Hall bar devices is currently limited by the precise controllability of the stray field, such that the required energy equilibrium for field-free SOT is not disturbed by the stray field. The Hallbar synapse, along with stochastic MTJ neurons, establishes essential hardware components for all-spintronic RBM and DBN circuits. While managing stray fields in full MTJ stacks is actively being optimized. Collectively, the synaptic properties, stochastic neuronal activation, and inherent sampling characteristics provide a scalable framework for all-spintronic neuromorphic systems, enabling classification, optimization, and other applications.

## VI. METHODS

### A. Device Fabrication

For Hall bar devices, a stack of Ta (5 nm)/Co$_{40}$Fe$_{40}$B$_{20}$ (0.65 nm)/MgO (2 nm)/Ta (2 nm) was deposited. Additionally, a stack of Co$_{40}$Fe$_{40}$B$_{20}$ (0.9 nm)/MgO (2 nm)/Ta (2 nm) was deposited as a reference sample for ion beam etching (described below). For MTJ devices, a multilayer stack consisting of Ta (5 nm)/Co$_{40}$Fe$_{40}$B$_{20}$ (0.65 nm)/MgO (2 nm)/Co$_{40}$Fe$_{40}$B$_{20}$ (1.3 nm)/Ta (0.6 nm)/Co (0.6 nm)/Pt (1.5 nm)/[Co (0.4 nm)/Pt (0.2 nm)]$_4$/Co (0.5 nm)/Ru
(1.0 nm)/Co (0.5 nm)/[Pt (0.2 nm)/Co (0.4 nm)]$_9$/Ta (8 nm) was deposited. Furthermore, a similar stack with Co$_{40}$Fe$_{40}$B$_{20}$ (0.9 nm) as the bottom magnetic layer was deposited as a reference sample for ion beam etching of the MTJ pillars.

The MgO layers were deposited by RF magnetron sputtering, while all other layers were deposited using DC magnetron sputtering. All films were deposited on Si/SiO$_2$ (300 nm) substrates using a ROTARIS sputter at room temperature, with a base pressure of approximately $2.2 \times 10^{-8}$ mbar.

#### 1. Hall Bar Device Fabrication

Before patterning into Hall bar devices, the films were annealed at 300 °C for 1 hour in vacuum ($\sim 5 \times 10^{-5}$

Pa) with an out-of-plane magnetic field of 8 kOe applied. Lithography was performed using a Heidelberg DWL66+ Maskless laser lithography system with AZ 5214E positive photoresist. Ar ion etching was carried out using an ADVANCEDMEMS ion beam etcher.

Initially, Hall bar patterns (with and without notches) were transferred onto the films using dark mode exposure, followed by development and ion beam etching to form standard Hall bars (with and without notches). In a second lithography step, the current channels of the Hall bars were covered with photoresist. Both the Hall bars and reference sample were etched simultaneously until the reference sample became insulating, thus removing the $Co_{40}Fe_{40}B_{20}$ (0.65 nm)/MgO (2 nm)/Ta (2 nm) stack from the Hall arms. A third lithography was conducted to define electrical contact pads, followed by deposition of Ti (10 nm)/Au (100 nm) using an ESC sputter system and a lift-off process.

### 2. MTJ Fabrication

The fabrication of magnetic tunnel junctions (MTJs) involved precise device design, incorporating alignment marks to enable accurate overlay between maskless laser lithography and electron beam lithography (EBL), achieving ∼ 100 nm resolution while minimizing exposure time. The patterning was performed using the Heidelberg DWL66+ maskless laser lithography system and the JEOL JBX-6300FS EBL system. The AZ 5214E photoresist was employed for laser lithography, while mAN-3403, a negative-tone resist, was used for high-resolution EBL patterning.

After dark-mode laser lithography and development, pattern transfer was defined via argon ion beam etching using an ADVANCEDMEMS ion beam etcher. Subsequently, nanoscale pillar structures were patterned by exposing the 300 nm-thick mAN-2403 resist at a dose of 300 $\mu C/cm^2$ using the JEOL EBL system operated at 100 kV and 100 pA. The EBL-processed samples, along with a reference sample, were etched simultaneously until the reference sample became electrically insulating, indicating the complete removal of the multilayer stack (excluding the Ta layer) from unprotected regions.

Without stripping the resist, the samples were immediately transferred to an AJA magnetron sputtering system for deposition of a 60 nm $SiO_2$ insulating encapsulation layer. The remaining resist was then removed via ultrasonic cleaning in N-Methyl-2-Pyrrolidone (NMP), resulting in well-defined openings atop the nanopillars for access to the top electrode.

Further, clear-mode laser lithography was used to define openings in the bottom channel electrode area. These were then etched via reactive ion etching (RIE) using the Oxford PlasmaLab System 100 ICP-RIE system to remove the $SiO_2$ at the designated contact sites. Finally, electrical contact pads were patterned using clear-mode maskless laser lithography, followed by Ti (10 nm)/Au (100 nm) deposition via ESC sputtering and a lift-off process, completing the MTJ device fabrication.

The MTJ devices were annealed at 300 °C for 1 hour in vacuum (∼ $5 \times 10^{-5}$ Pa) with an out-of-plane magnetic field of 8 kOe applied.

### B. Electrical Measurements

Electrical measurements were performed using an Eastman Kodak EM3P system, equipped with two Keithley Model 6221 current sources and a Keithley Model 2182A nanovoltmeter.

For multi-pulse measurements, one Keithley 6221 was used to apply current pulses with a typical pulse width of 10 $\mu$s (unless otherwise specified) to serve as writing currents, while the other Keithley 6221 provided a small DC reading current (typically 0.1 mA) a few seconds after the writing current pulse. The Hall voltage under the reading current was measured using the Keithley 2182A nanovoltmeter.

For MTJ neuron characterization (time-dependent $R_{MTJ}$ measurements), one Keithley 6221 was used to apply a relatively large DC through the bottom channel of the MTJ. A 10 MΩ resistor was connected between ground and the MTJ top electrode, such that the resistor was in series with the MTJ and allowed approximately 0.01% of the DC to tunnel through the MTJ and flow through the resistor. The resistance state of the MTJ was determined by measuring the voltage across the 10 MΩ resistor using the Keithley 2182A nanovoltmeter. The MTJ resistance ($R_{MTJ}$) was calculated using the following formula:

$$R_{MTJ} = \frac{1}{2} R_B \left( \frac{R_{\text{ref}}}{V_{\text{ref}}} I_{DC} - 1 \right) - R_{\text{ref}} \tag{14}$$

where $R_B$ is the resistance of the MTJ bottom channel, $R_{\text{ref}}$ is the resistance of the reference resistor (10 MΩ), $V_{\text{ref}}$ is the measured voltage across the resistor, and $I_{DC}$ is the applied DC.

### C. MOKE Measurements

MOKE measurements were conducted using a TuoTuo Technology TTT-02 Kerr Microscope System, equipped with a Keithley Model 6221 current source and a Keithley Model 2182A nanovoltmeter. A red laser with a wavelength of 635 nm was utilized as the imaging light source during the measurements. The Keithley 6221 was used to apply current pulses with a typical pulse width of 10 μs (unless otherwise specified), serving as writing currents. Initially, a background image was captured when the Hall bar device was in a saturated magnetized state, achieved by applying a large current pulse (8 mA) in conjunction with an in-plane magnetic field of 26 Oe. Subsequently, after each current pulse, a differential image was recorded using the Kerr microscope system. When applicable, an external magnetic field was applied, and its magnitude was measured using an external Gaussmeter.

### D. Modelling and Simulations

The total magnetic energy of the 0.65 nm *CoFeB* magnetic layer used in the simulations includes exchange, Zeeman, uniaxial anisotropy, demagnetization, and DMI energies.

$$E(\boldsymbol{m}) = \int_V [A(\nabla \boldsymbol{m})^2 - \mu_0 \boldsymbol{m} \cdot H_{\text{ext}} - \frac{\mu_0}{2} \boldsymbol{m} \cdot H_d - K_u(\hat{u} \cdot \boldsymbol{m}) + \varepsilon_{\text{DM}}] dv \tag{15}$$

where $A$ is the exchange stiffness, $\mu_0$ is the permeability, $K_u$ is the anisotropy energy density, $H_d$ is the demagnetization field, and $H_{ext}$ is the external field; moreover, the DMI energy density is then computed as follows:

$$\varepsilon_{\text{DM}} = D[m_z(\nabla \cdot m) - (m \cdot \nabla) \cdot m] \tag{16}$$

Micromagnetic simulations were conducted with MuMax [53], which employs the Landau–Lipschitz–Gilbert (LLG) equation as the core principle for calculating magnetization dynamics. The LLG equation captures the magnetization dynamics in the following manner:

$$d\boldsymbol{m}/dt = -\gamma/((1 + \alpha^2)) [m \times H_{eff} + \alpha m \times (m \times H_{eff})] + \tau SOT$$

$$\tau SOT = -(-\gamma/((1 + \alpha^2))) aJ [(1 + \xi\alpha)m \times (m \times p) + (\xi - \alpha)(m \times p)] \tag{17}$$

Where $\boldsymbol{m}$ is the normalized magnetization vector, $\gamma$ is the gyromagnetic ratio, $\alpha$ is the Gilbert damping coefficient, and

$H_{\text{eff}}$ is the effective magnetic field around which the magnetization process occurs. The spin–orbit torque is then added as modified STT in $a_J = \left| \frac{\hbar}{2M_S e \mu_0} \frac{\theta_{\text{SH}} j}{d} \right|$ MuMax.

and $p = sign(\theta_{\text{SH}})j \times n$ Where $\theta_{\text{SH}}$ is the spin Hall coefficient of the material, $j$ is the current density, and $d$ is the free layer thickness.

The simulation parameters are given in Tab. 2 below. We defined the 512x128 nm² tracks with side arms, as shown.

Table I. Magnetic simulation details

| Grid Size | Cell Size (nm) | Anisotropy Ku ($\frac{J}{m^3}$) | Saturation Mag.Ms ($\frac{A}{m}$) | Exchange Stiffness ($\frac{J}{m}$) | DMI $J/m^2$ |
|---|---|---|---|---|---|
| 512, 128, 1 | $\frac{150}{128}, \frac{150}{128}$, 0.65 | $0.9 \times 10^6$ | $0.8 \times 10^6$ | $1.6 \times 10^{-11} \times 10^{-3}$ | $0.2 \times 10^{-3}$ |

In (SF1), which matches the designs of asymmetric and notched devices. Multiple batch simulations were conducted with increasing current densities in the absence of any magnetic field. The micromagnetic images and videos have been added to SF1.

## ACKNOWLEDGMENTS

The authors thank Dr. Nabeel Aslam and Dr. Naveed Riaz Kazmi for their support during the device fabrication.

### E. Author Contribution

A. H. L. conceived the idea of a field-free SOT spintronic synapse device and stochastic MTJs, as well as their integration in RBM. A. H. L., M. T., and C. F. planned the device designs. M. T. and C. F. fabricated the devices. M. T. and A. H. L. planned and performed the electrical measurements of the synapse and neuron devices. M. T., B. H, and J. X. performed the MOKE imaging. A. H. L. conducted the micromagnetic simulations, synapse, neuron modeling, and their integration in the RBM architecture. A. H. L. wrote the manuscript with support from M. Tang, C. F, and G. S.. All authors participated in the discussion of the results and review. G. S. provided supervision throughout the project.

### F. Data Availability

The data associated with this study are available from the authors upon reasonable request.